# A Call to Promote Soft Skills in Software Engineering

Luiz Fernando Capretz, Ph.D., P.Eng.
Western University
Dept. Electrical & Computer Engineering
London, Ontario, Canada – N6A5B9
lcapretz@uwo.ca

Faheem Ahmed, Ph.D., P.Eng.
Thompson Rivers University
Dept. Computing Science
Kamloops, BC, Canada – V2C0C8
fahmed@tru.ca

We have been thinking about other aspects of software engineering for many years; the missing link in engineering software is the soft skills set, essential in the software development process. Although soft skills are among the most important aspects in the creation of software, they are often overlooked by educators and practitioners. One of the main reasons for the oversight is that soft skills are usually related to social and personality factors, i.e., teamwork, motivation, commitment, leadership, multi-culturalism, emotions, interpersonal skills, etc. This editorial is a manifesto declaring the importance of soft skills in software engineering with the intention to draw professionals' attention to these topics. We have approached this issue by mentioning what we know about the field [1-6], what we believe to be evident [7-9], and which topics need further investigation [10-14]. Important references to back up our claims are also included.

Software engineers take pride in the depth of their technical expertise, which separates them from the crowd. But, what makes a good software engineer? First, it is the technical knowledge of relevant methodologies and techniques (i.e. hard skills), as well as the skills necessary for applying that knowledge in practice. Second, but nonetheless important, it is a set of soft skills, in particular collaboration, communication, problem-solving and similar interpersonal and critical thinking skills that are expected from software engineering professionals. In other words, software engineers need both hard and soft skills in order to be successful at the workplace [15].

Engineering software involves performing tasks in distinct areas, such as system analysis, software design, programming, software testing, and software evolution/maintenance; other software occupations in a software team include the project manager, troubleshooter, helpdesk personnel, database administrator, and so forth. Thus today, specialties within software engineering are as diverse as in any other profession [16]. Additionally, software engineers need to communicate very effectively with users and team members, which reinforces the idea that the people dimension of software engineering is as important as the technical expertise [17].

However, computer science and software engineering curricula focus mainly on developing hard skills, thus paying lip services to soft skills. Even the latest guideline for teaching and learning software engineering, namely the SWEBOK V3.0 and IEEE/ACM Curriculum Guide, praises technical competence and gives marginal consideration to vaguely characterized non-technical (soft) skills. We strongly believe that computer science and software engineering curricula should put more emphasis on developing and assessing both hard and soft skills, and that both types of skills should be acknowledged.

Psychology tells us that not everybody excels at all types of tasks. Successful software development also depends significantly on how software practitioners perform their tasks and how they interact with their peers.

At present time, very few courses in computer science or software engineering curricula touch upon the subjects of teamwork and soft skills assessment [18]. It is difficult even to find a university that has an entire course on the human aspects of software engineering, with the exception of a couple of computer science departments [19-20]. Unfortunately, soft skills are far from being a part of mainstream software engineering education; a course devoted to this subject would be an ideal solution to remedy this situation. It is important to notice here that this is not only our opinion; when we met senior software engineers who were leading teams, they told us that there was a need for a soft skills set among software developers. The productivity and satisfaction of technical people have been seriously undermined by the lack of this kind of knowledge in computer science education, software engineering, and related areas.

Engineering software revolves around methodologies establishing (more or less) well-defined processes to develop a product [21]. However, because people execute these processes, no matter how good they are, the ultimate success lies in how the processes are being executed especially when people interact. The successful implementation of any process methodology ultimately depends on how employees perceive that methodology. For example, besides the criticism, iterative and incremental approaches of Agile methodology have paved the way to reduce costs and shorten development time. Arguably, Agile is not a methodology but a mindset or a different way of managing people to develop software products. Although Agile is suited for many practitioners it is not suited for everyone [22]. Agile methodologies emphasize on some social aspects of the process such as competency, collaboration, trust, analytics, and devolution of decision-making. However, numerous other social attributes which vary from organization to organization such as work ethics, respectful work environment, reward mechanism, internal politics, relationship management, conflict management, etc. are not addressed and have been given little or no attention by the software industry in general.

Furthermore, software engineers who are promoted to managerial positions, typically experiment similar emotions like others do. In addition to earning more money, and possibly a better parking spot, promotions are also a recognition of their expertise and dedication by their superiors and peers. At the same time, promotions also cause headaches for software engineers. In addition to the common challenges, such as becoming the boss of their previous colleagues (or foes), software managers need to manage conflicts, negotiate and compromise, measure performance, etc. Software engineers often struggle to adopt a managerial culture because they are not trained to be managers. Thus, software engineers moving from a technical to a managerial position struggle with the transition because now their focus shifts from product development to managing people, which is relatively more difficult.

The success or failure of a software manager resides heavily on the performance and creations of others. The dilemma of having technical knowledge and a managerial position at the same time often forces them to wade in and attempt to fix problems that rightly belong to a subordinate. Software managers may also find it hard to accept that, eventually, someone younger than them and working under them will outstrip their technical knowledge of some domain because they themselves are too busy managing people to keep up with rapid changes in technology. One of the reasons why software managers tend to keep looking at the trees instead of at the forest (that is, the big picture) is that they have been trained under the notion of the predictable behavior of software, whereas the behavior of people is less predictable.

In summary, technical people tend to overlook the importance of soft skills as it is unrelated to their technical area and because their training is in dealing with technical issue; thus considering the soft skills in the software development process to be foreign to them, since the field deals with human factors and touches social sciences. These are topics that software professionals do not have expertise in. We believe that it is high time for the software development community to realize that the human element is pivotal to success in the engineering of software. We have to recognize that software engineering is a people-intensive discipline, hence requires appropriate treatment. Therefore, human aspects of software engineering are important subjects to teach, study and research. We urge software engineers to take on this challenge.

**Important References:**

**About the Authors:**

**Luiz Fernando Capretz** is a professor of software engineering and assistant dean (IT & e-Learning) at Western University in Canada, where he also directed a fully accredited software engineering program. He has vast experience in the engineering of software and is a licensed professional engineer in Ontario. Contact him at lcapretz@uwo.ca or via www.eng.uwo.ca/people/lcapretz .

**Faheem Ahmed** is a professor and Chair in the Department of Computer Science at Thompson Rivers University in Kamloops, British Columbia, Canada. His research interests are software process assessment, empirical software engineering, and green computing. He can be reached at fahmed@tru.ca.